\documentclass[12pt]{iopart}

\usepackage{graphicx}%
\usepackage{multirow}%
\usepackage{amssymb,amsfonts}%
\usepackage{amsthm}%
\usepackage{mathrsfs}%
\usepackage[title]{appendix}%
\usepackage{xcolor,xspace}%
\usepackage{textcomp}%
\usepackage{manyfoot}%
\usepackage{booktabs}%
\usepackage{algorithm}%
\usepackage{algorithmicx}%
\usepackage{algpseudocode}%
\usepackage{listings}%
\usepackage{siunitx}
\usepackage{amssymb}
\usepackage[noabbrev]{cleveref}
\usepackage{url} 
\usepackage{paralist}
\usepackage[shortlabels]{enumitem}

\newcommand{\kindex}[2]{\ensuremath{{#1}_{\scalebox{0.65}{#2}}}}

\newcommand{\Uinf}{\kindex{U}{$\infty$}\xspace}
\newcommand{\dd}[1]{\textrm{d}\ensuremath{#1}\xspace}

\definecolor{matlabblue}{rgb}{0, 0.4470, 0.7410}
\definecolor{matlaborange}{rgb}{0.8500, 0.3250, 0.0980}
\definecolor{chocolate}{rgb}{0.8203, 0.4102, 0.1172}
\definecolor{canard}{rgb}{0, 0.4531, 0.5000}
\definecolor{black}{rgb}{0, 0, 0}

\begin{document}

\title{Event-based reconstruction of time-resolved centreline deformation of flapping flags}

\author{Gaétan Raynaud \& Karen Mulleners}

\address{{Unsteady Flow Diagnostics Laboratory, Institute of Mechanical Engineering, École Polytechnique Fédérale de Lausanne (EPFL), Lausanne, 1015, Switzerland}}
\ead{karen.mulleners@epfl.ch}
\vspace{10pt}
\begin{indented}
	\item[]March 2024
\end{indented}








\begin{abstract}
High-speed imaging is central to the experimental investigation of fast phenomena, like flapping flags. 
Event-based cameras use new types of sensors that address typical challenges such as low illumination conditions, large data transfer, and the trade-off between increasing repetition rate and measurement duration more efficiently and at reduced costs compared to classical frame-based fast cameras. 
Event-based cameras output unstructured data that frame-based algorithms can not process. 
This paper proposes a general method to reconstruct the motion of a slender object similar to the centreline of a flapping flag from raw streams of event data. 
The method takes advantage of continuous illumination, and the reconstruction update rate is set after and independent of the data collection.
Our algorithm relies on a coarse chain-like structure that encodes the current state of the line and is updated by the occurrence of new events. 
The algorithm is applied to synthetic data, generated from known motions, to demonstrate that the method is accurate up to one percent of error for tip-based, shape-based, and modal decomposition metrics. 
Degradation of the reconstruction accuracy due to simulated defects only occurs when the severity of the defects is more than two orders of magnitude larger than what we typically encounter in experiments.
The algorithm is then applied to experimental data of flapping flags, and we obtain relative errors below one percent when comparing the results with the data from laser distance sensors. 
The reconstruction of line deformation from event-based data is accurate and robust, and unlocks the ability to perform autonomous measurements in experimental mechanics.
\end{abstract}



\vspace{2pc}
\submitto{\MST}
%
\maketitle
%

\section{Introduction}\label{sec1}

\begin{figure}
\centering
\includegraphics[width=\linewidth]{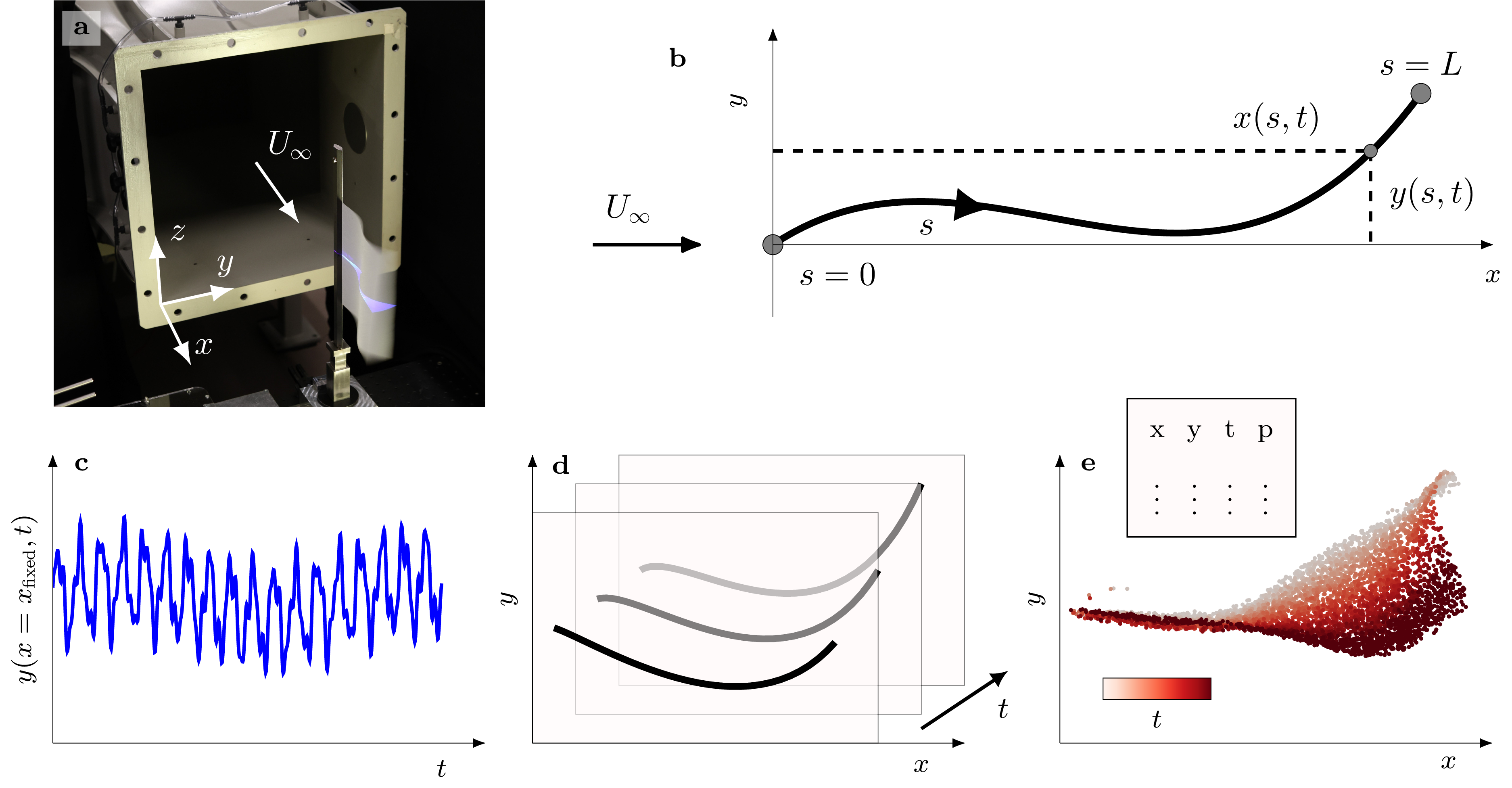}
\caption{
Example of the measurement of the centreline of a flapping flag. 
(a) Photograph of the flapping flag at the wind tunnel outlet. 
The 2D motion of the centreline is lit up in purple and recorded with the event-based camera. 
(b) Coordinates $x(s,t)$ and $y(s,t)$ describe the 2D motion of the centreline along the curvilinear abscissa $s$  at time $t$. 
The root ($s=0$) is fixed and the tip ($s=L$) is free to move. 
The free-stream flows along the $x$-axis with velocity \Uinf. 
The information used to reconstruct the centreline motion can be formatted as (c) a 1D transverse displacement at a fixed downstream position $x=\kindex{x}{fixed}$, (d) a series of instantaneous snapshots of the illuminated centreline or (e) a table of individual events plotted as a cloud coloured by time.}
\label{fig:CasePresentation}
\end{figure}

Many phenomena in experimental mechanics are fast and require fast imaging techniques.
Dynamic crack propagation \cite{freund_dynamic_1998}, buckling rods \cite{gladden_dynamic_2005}, snap-through beams \cite{gomez_critical_2017}, or fluttering foils are fast phenomena that are characterised by time scales of the order of microsecond to millisecond.
Flags and slender flexible structures that are immersed in a flow are subject to flutter, a dynamic instability \cite{yu_review_2019}. 
Experimental studies of flags in a wind tunnel help improve our understanding and modelling of unsteady aerodynamics, with applications in energy harvesting \cite{eugeni_numerical_2020}, biological tissues \cite{auregan_snoring_1995,johnson_effects_2022} and bio-inspired propulsion \cite{muller_fish_2003}.
Fluttering flag experiments in a wind tunnel (\cref{fig:CasePresentation}.a) involve flapping frequencies \kindex{f}{flapping} from \qtyrange{10}{50}{\Hz} \cite{virot_fluttering_2013}.
The flapping behaviour of flags is characterised by their time-resolved deformation for various wind speeds $\Uinf$ (\cref{fig:CasePresentation}.b).

Measuring the deformation of flapping flags with high temporal and spatial resolution is challenging and costly.
Distance sensors can provide displacement signals with high temporal resolution but only at a single spatial location (\cref{fig:CasePresentation}.c).
Frame-based imaging captures space-resolved data at subsequent time instants, but save irrelevant or repeated information for a large portion of the pixels in the snapshots (\cref{fig:CasePresentation}.d).
Cameras with a higher spatial resolution can typically record for a long time at low frame rates. 
Cameras with a higher frame rate using modern global shutter sensors are associated with lower spatial resolution and shorter recording times as the amount of data is limited by the transfer rate and size of the camera buffer \cite{raffel_recording_2018}.
Such high frame rate cameras are expensive, and the associated high shutter speeds call for higher intensity illumination to avoid the signal-to-noise ratio to drop.
Event-based cameras continuously record dynamic variations of light at the pixel level \cite{gallego_event-based_2022}, creating a flow of pixel-level information in the moving areas only (\cref{fig:CasePresentation}.e).
Event-based cameras unlock new capabilities for high spatio-temporal resolution tracking of objects, with no motion blur, and high dynamic range, at a fraction of the price \cite{gallego_event-based_2022}.

The difference in the data structure between event-based and frame-based imaging calls for a paradigm shift in the development of processing algorithms.
Specific applications include the tracking of objects \cite{litzenberger_embedded_2006,ghosh_real-time_2014,conradt_pencil_2009}, micro particles \cite{ni_asynchronous_2012}, or seeding particles for flow field measurements using particle tracking or particle image velocimetry \cite{drazen_toward_2011,willert_event-based_2022-1,willert_event-based_2022}. 
Recent contributions to event-based extraction of the centreline deformation of flexible membranes \cite{lyu_high-frequency_2024} and seeding particles \cite{willert_event-based_2022} still relied on pulsed illumination to generate discontinuous bursts of events.
Raw data are sampled at each light pulse and processed one pulse at a time. 
Pulsed illumination compels the user to know a priori the typical time scales of the recorded phenomena, and duplicates static information at each pulse.
Here, we present an algorithm that uses continuous illumination and leverages the full potential of event-based imaging to reconstruct the centreline deformations of a flapping flag with arbitrary time discretisation.

\begin{figure}
\centering
\includegraphics[width=\linewidth]{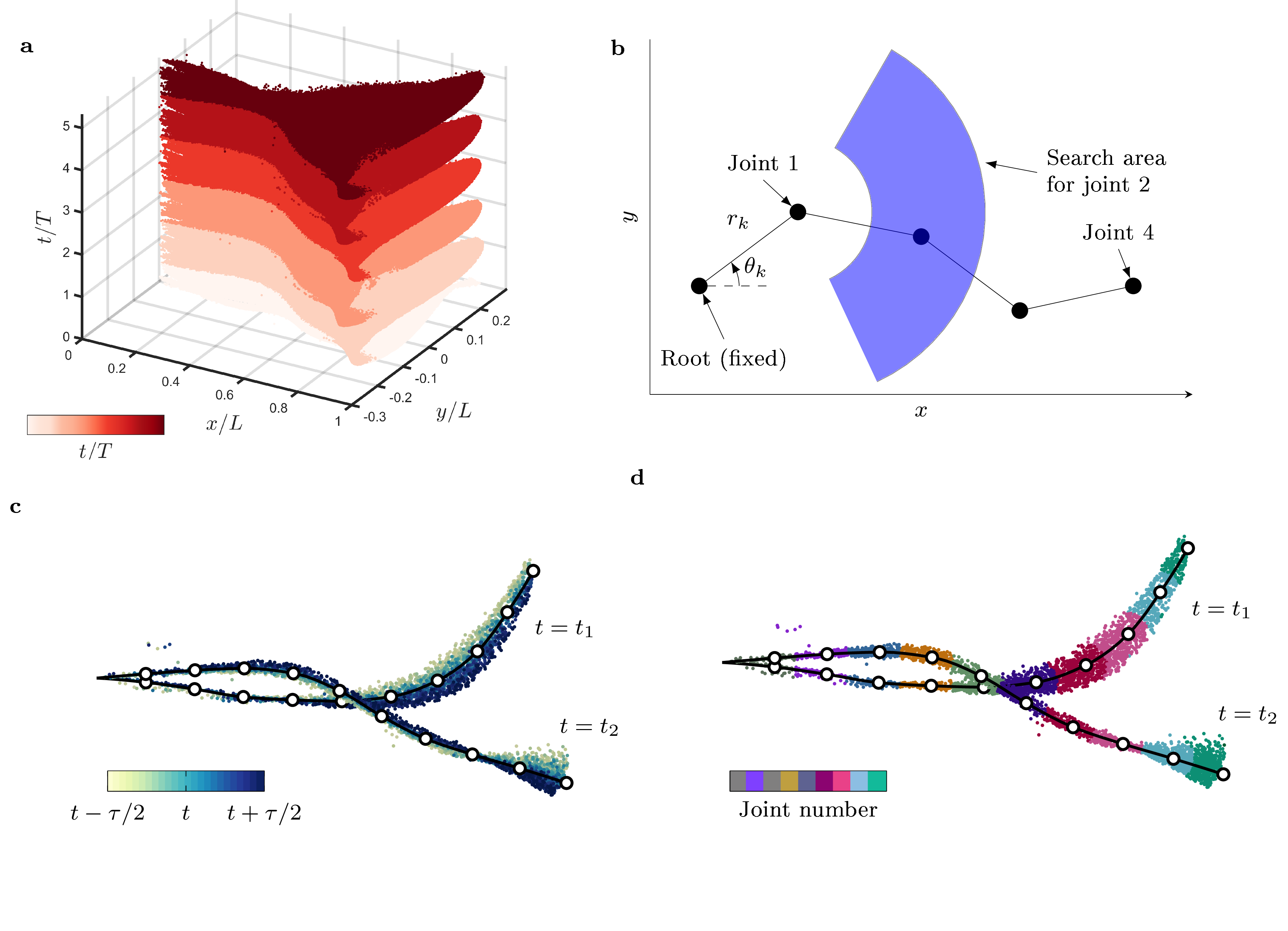}
\caption{(a) The event-based camera outputs a cloud of events following the shape of the centreline motion in the $(x,y,t)$ space. 
(b) The coarse structure that encodes the current state of the centreline consists of links and joints. 
The relative location of joint \kindex{r}{k} with respect to joint \kindex{r}{k-1} is given by the polar coordinates~$(\kindex{r}{k},\kindex{\theta}{k})$.
The coloured arc represents the search area around the current joint position to update the coordinates in the next time step. 
(c) Two blocs of events are projected on the $(x,y)$ plane and overlaid with the reconstructed line and joints positions. 
The colour indicates the time of the events for two blocks of duration $\tau/T = \num{6.2e-2}$ at different moments in time. 
(d) The same events as in (c) are shown, coloured by the search area surrounding individual joints.}
\label{fig:algorithm_illustration}
\end{figure}

\section{Method}
\label{sec:method}
Event-based cameras record changes in intensity at the pixel level.
Here, we consider a flapping flag in a uniform flow that is imaged using an event-based camera to measure the centreline deformation of the flag in time (\cref{fig:CasePresentation}).
The centreline is illuminated by a laser sheet normal to the flag that intersects the flag at mid-height. 
The laser used is a continuous laser to leverages the full potential of event-based imaging.
An event-based camera records the reflections of the laser sheet on the flag from above. 
If the flag is stable and remains still, the centreline does not move in the field of view.
The light intensity at each pixel stays constant over time and no events are generated.
When the flag flaps, the position of the illuminated centreline moves in the field of view.
Photoreceptors undergo light intensity transitions when the centreline reflections pass them.
Individual pixel events are generated at the precise time and location in the field of view corresponding to the locations where the centreline passes by.
The absence of events for a still object in the field of view differs from studies using pulsed illumination.
When pulsed illumination is used, events are generated for every pulse, whether the object's centreline moves or not.
The list of events can be represented as a cloud indicating the changes in centreline deformation in space and time (\cref{fig:algorithm_illustration}.a). 
Each event is described by a spatial position $\kindex{x}{evs}$ and $\kindex{y}{evs}$, a time stamp $\kindex{t}{evs}$ and a polarity $\kindex{p}{evs} = 1$ or $0$ for increasing or decreasing brightness, respectively. 
The spatial coordinates obtained after calibration (see Appendix 1) are non-dimensionalized by the centreline's length $L$ and time stamps $\kindex{t}{evs}$ is non-dimensionalized by the flapping period $T=1/\kindex{f}{flapping}$ which is determined a posteriori.
This data structure is spatially sparse and provides only information about changes in the field of view. 
Each event indicates where something is moving, but it does not tell us which part of the flag is moving and where the rest of the flag is. 

A specialised algorithm is required to convert the cloud of events into a parametrised description of the deformation of the flag's centreline expressed by the coordinates $x(s,t)$ and $y(s,t)$ with $s$ the curvilinear abscissa and $t$ the time.
The solution we propose here is an algorithm that encodes the current state of the geometry in time with a chain-like structure of $\kindex{N}{j}$ links and joints, excluding the fixed root connection.
The relative location of successive joints is defined in polar coordinates~$(\kindex{r}{k},\kindex{\theta}{k})$ (\cref{fig:algorithm_illustration}.b).
The total length of the chain $L = \sum_{k = 1}^{\kindex{N}{j}} \kindex{r}{k}$ is either kept constant for inextensible objects, or might vary in time if the material is extensible.
Here, we will focus solely on inextensible flags, but the algorithm can be extended in future work to account for local stretching or compression of the material as a function of known material properties and additional measures of the local shear stress, for example. 
The initial distribution of the joints along the object is chosen to be equidistant here. 
Alternative non-equidistant distributions can be selected if desirable.  
The variations of $\kindex{\theta}{k}$ along the chain describe the line curvature.

To update the chain coordinates, we consider events during an update interval of $\tau$ centered around the current processing time $t$ (\cref{fig:algorithm_illustration}.c). 
Events with time stamps between $t-\tau/2$ and $t+\tau/2$ are considered to define the shape at time $t$.
These events are then spatially clustered starting from the root to successively update the location of the joints from the root to the tip of the flag (\cref{fig:algorithm_illustration}.d). 
For each joint $k$, we define an arc-shaped spatial search area as indicated in \cref{fig:algorithm_illustration}.b. 
Events are within the search area if their distance from the previous joint $k-1$ lies between $(1-\alpha)\kindex{r}{k}$ and $(1+\alpha)\kindex{r}{k}$ with $\alpha$ the search radius parameter, which is a scalar value between \numrange[range-phrase={\ and\ }]{0}{1} ($\alpha=0.6$ in \cref{fig:algorithm_illustration}.d).
The updated location of joint $k$, \kindex{\theta}{k}, is then determined by the average angle with respect to joint $k-1$ of all the events with a timestamp within the update interval and a spatial location in the search area. 
The search and update process is repeated for all successive joints up to $k = \kindex{N}{j}$ for a selected time instant $t$ to obtain $(\kindex{r}{k}, \kindex{\theta}{k})$ pairs for each link. 

This iterative algorithm is robust to spatial and temporal discontinuities in the recording of event clouds.
The most common reasons for discontinuities in the recording of event clouds are non-moving parts of the structure, e.g. near the root of the flag, or interruptions in the light source, e.g. brief flashing of the continuous light source or shadow on the camera lens.  
If no events are found in a search area for a specific update interval due to the absence of motion, the coordinates of the corresponding joint are not updated, respecting the stationarity of the structure. 
A short-term failure or interruption of the light source or blockage of the camera lenses does not perturb the continuation of the iterative processing algorithm, as the full structure will generate events again once the interruption or blockage are resolved. 

The joint coordinate pairs $(\kindex{r}{k}, \kindex{\theta}{k})$ are converted to Cartesian coordinates $(\kindex{x}{k},\kindex{y}{k})$. 
To smoothen this discrete joint-based centreline deformation, we replaced the straight links between joints by piece-wise polynomials using the modified Akima cubic Hermite interpolation in Matlab with a final spatial curvilinear discretisation of $\dd{s}$.
The interpolation increases the length of the centreline beyond the flag length. 
This is corrected for by reducing the radial coordinate $\kindex{r}{\kindex{N}{j}}$ of the tip joint and maintaining the extracted tip angle \kindex{\theta}{\kindex{N}{j}} until the length of the smoothened centreline equals the flag length.
This entire process is repeated with a time increment $\dd{t}$ as long as there are new events in the stream.

The main novelty of our approach is the use of a chain-like structure that is updated iteratively for the extraction of the spatiotemporal evolution of the centreline of a deforming structure from a cloud of events.
The proposed algorithm preserves the length of the structure and only updates the joint coordinates when motion is detected through new events. 
A local displacement of the tip will thus not affect the reconstruction at the root of the flag if no motion is detected there. 
The reconstruction update rate of the joint coordinates is arbitrarily set by the user after the data is gathered and does not need to be constant in time.
This property makes it possible to record long time series of transient behaviour covering faster and slower motions in a single file without any risk of under- or oversampling since the temporal discretisation of the deformation extraction is set and adapted during processing based on the recorded data.
Our approach is therefore more intuitive and robust than a direct fitting of the event clouds to obtain the centreline coordinates. 
Curvilinear coordinates can not be directly obtained using fitting as the curvilinear coordinate $s$ associated with an event is not known a priori, which is especially challenging for curves with large displacements. 
For approaches based on direct fitting, it is also more difficult to implement constraints such as preserving a fixed overall length and to guarantee that local displacements, e.g. at the tip, do not affect the entire shape.

The algorithm presented here uses five main user-input parameters: the number of joints $\kindex{N}{j}$, the output spatial resolution $\dd{s}$, the search radius parameter $\alpha$, the processing time increment $\dd{t}$, and the update interval $\tau$.
The number of joints $\kindex{N}{j}$ defines the coarseness of the chain and controls the resolution of the local curvature and the spatial deformation wavelength.
The number of joints should be approximately an order of magnitude larger than the number of spatial deformation wavelengths that fit in the flag length and should be further increased with increasing local curvature. 
The spatial curvilinear resolution $\dd{s}$ controls the smoothness of the resulting centreline.
It should be equal to or smaller than the curvilinear distance between the joints.
The search radius parameter $\alpha$ controls the size of the search area and the local spatial smoothening of the deformation. 
Ideally $\alpha$ should be as small as possible to reduce smoothing, but large enough such that enough events are included to reduce the random error.
Values of $\alpha \geq 0.5$ generate overlap between the search areas of two successive joints.
The processing time increment $\dd{t}$ and the update interval $\tau$ determine the temporal discretisation. 
Ideally, the update interval is as short as possible to increase the accuracy of the local deformation velocity, but it needs to be large enough such that enough events are included to reduce the random error.
The processing time increment should preferably be smaller than the update interval ($\dd{t}<\tau$) to ensure continuous update of the chain-like structure in our algorithm as some events will be considered for multiple time instants and sudden jumps are avoided.
If there is no temporal overlap for values of $\dd{t}>\tau$, some events will be ignored which can lead to large jumps between position updates that will cause the algorithm to lose track and fail. 

The choice of the input parameters is a compromise between precision and robustness.
The influence of the processing parameters is further discussed in the next section using a benchmark test with simulated events.

\section{Benchmark of the algorithm using a simulated stream of events}
\label{sec:bemchmark}

In this section we test the robustness of the proposed method using synthetic data.
A total of \num{400} kinematics that are qualitatively similar to flapping flags are generated and exported in a video format.
These kinematics consist of a sum of a travelling wave and a standing wave with a growing amplitude envelope along the flag (supplementary video S1).
The frequency, wavelength and weight of both waves are randomly sampled to obtain a group of kinematics with peak-to-peak amplitudes $2A/L$ ranging from \numrange{0.085}{1.05} and wavelengths $\lambda/L$ ranging from \numrange{1.5}{3.0}.
The videos of the simulated motions are converted into streams of events using the function \verb|SimulatedEventsIterator| from Metavision SDK \cite{prophesee_metavision_2023}.
These video-to-events simulations estimate successive changes of intensity between frames to mimic the properties of an event-based camera and output a file containing a stream of events \cite{gehrig_video_2020,hu_v2e_2021}. 
We use our algorithm to reconstruct the deformation for a fixed duration of the recording corresponding to \numrange{3}{6} flapping periods using $\kindex{N}{j} = 10$ joints.

\begin{figure}
\centering
\includegraphics[width=\linewidth]{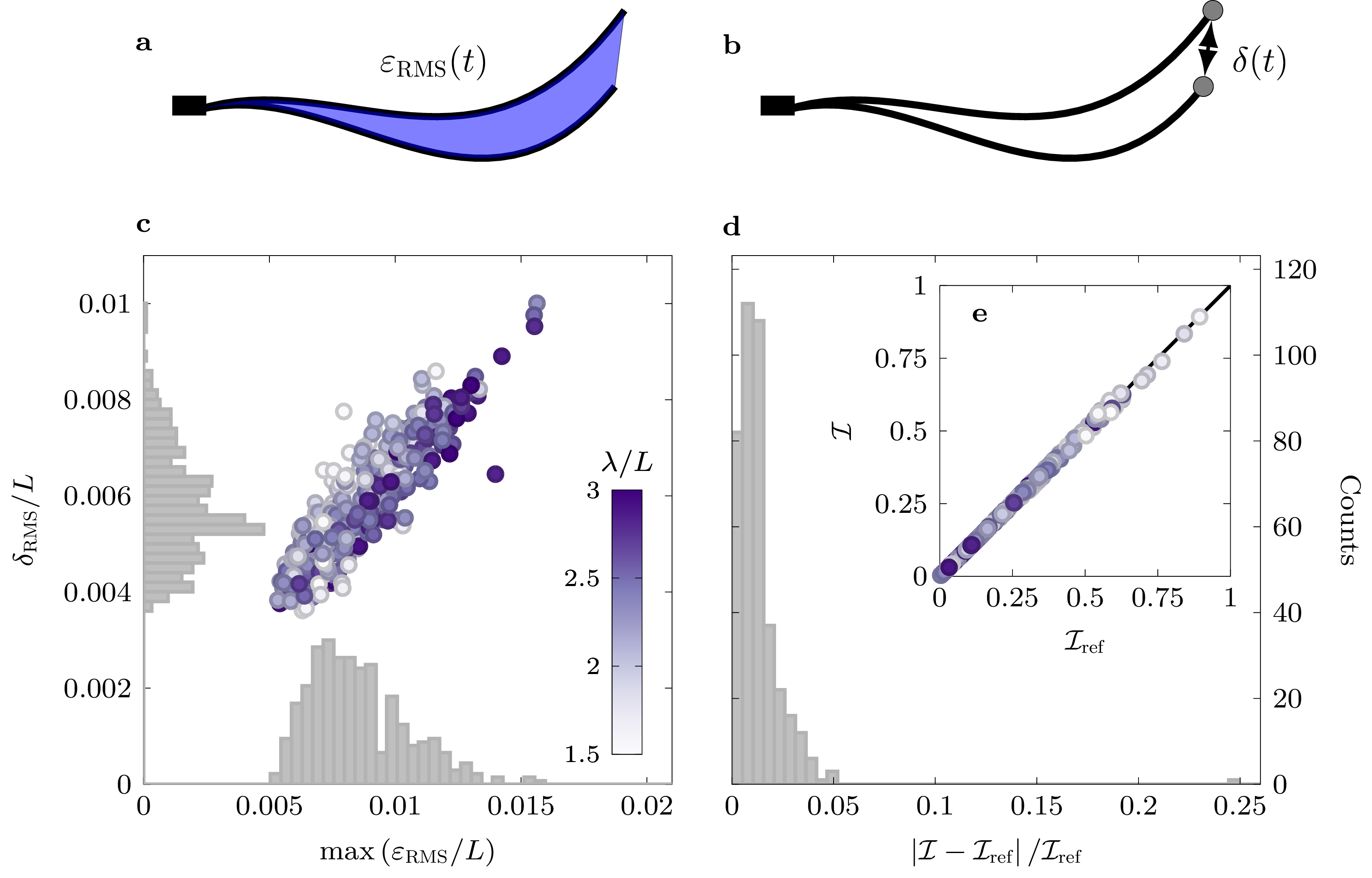}
\caption{
Quantification of the errors between the reconstructed and the exact deformation.
(a) Visualisation of the shape difference determined as the root-mean-square distance along the curvilinear abscissa $\kindex{\varepsilon}{RMS}(t)$, and (b) the tip displacement error $\delta(t)$.
(c) Distribution of root-mean-square tip displacement error $\kindex{\delta}{RMS}/L$ and maximum of the shape difference $\max\left( \kindex{\varepsilon}{RMS}/L\right)$ for various kinematics tested coloured by the dimensionless wavelength $\lambda/L$. 
Histograms are displayed for both quantities individually.
(d) Distribution of the relative travelling wave index error.
(e) Comparison of the reconstructed travelling wave index $\mathcal{I}$ with its reference value $\kindex{\mathcal{I}}{ref}$.}
\label{fig:kinematic_variations}
\end{figure}

The results of the comparison between the obtained shape $\left(x(s,t),y(s,t)\right)$  with the known input reference $\left(\kindex{x}{ref}(s,t),\kindex{y}{ref}(s,t)\right)$ are presented in \cref{fig:kinematic_variations}.
To evaluate the quality of the results, we propose different metrics to quantify the error in the reconstruction of the overall flag shape and the tip motion, and a modal-decomposition-based metric to compare the flapping dynamics. 
The instantaneous reconstruction error of the overall shape is calculated as the root-mean-square of the difference between the reconstructed and the reference position along the curvilinear abscissa (\cref{fig:kinematic_variations}.a):
\begin{equation}
\kindex{\varepsilon}{RMS} (t) = \sqrt{ \frac{1}{L}  \int\limits_{s=0}^{s=L} \left[x(s,t) - \kindex{x}{ref}(s,t)\right]^2 + \left[y(s,t) - \kindex{y}{ref}(s,t)\right]^2 \dd s }.
\end{equation}
The integral uses the smoothened piece-wise polynomial curve $\left(x(s,t),y(s,t)\right)$ with spatial discretisation $\dd s$ obtained by applying the modified Akima cubic Hermite interpolation of the Cartesian coordinates $(kindex{x}{k}, \kindex{y}{k})$ of the joints of the chain-like structure. 
For each reconstructed time-series we take the maximum of \kindex{\varepsilon}{RMS} over time as our shape-based metric to evaluate the quality of the instantaneous shape reconstruction. 
The displacement of tip of the flag plays an important role in the wake formation and the characterisation of the flapping dynamics. 
Therefore, we also consider a tip-based metric to evaluate our algorithm. 
As a tip-based metric we use the temporal root-mean-square value of the Euclidean distance $\delta(t)$ between the reconstructed location and the reference location of the tip (\cref{fig:kinematic_variations}.b): 
\begin{equation}
\delta(t) = \sqrt{\left[x(L,t) - \kindex{x}{ref}(L,t)\right]^2 + \left[y(L,t) - \kindex{y}{ref}(L,t)\right]^2}\quad.
\end{equation} 
The final metric we use to evaluate our algorithm's performance to reconstruct the flapping flag's dynamics is the travelling wave index $\mathcal{I}$.  
The travelling wave index indicates the relative contribution of travelling waves versus standing waves to the dynamics of a flapping motion \cite{feeny_complex_2013}. 
If the flapping motion is a pure standing wave, $\mathcal{I} = 0$, if the motion is a pure travelling wave, $\mathcal{I} = 1$. 
The travelling wave index is obtained using the complex orthogonal decomposition of the spatio-temporal evolution of the flag's shape. 
Matching travelling wave indices between the reconstructed deformations and the input shapes indicate a reliable reconstruction of the flapping dynamics.

For all motions considered, the relative root-mean square tip displacement error $\kindex{\delta}{RMS}/L$ ranges from \qtyrange{0.36}{1.0}{\percent} and the maximum shape difference $\max(\kindex{\varepsilon}{RMS}/L)$ ranges from \qtyrange{0.53}{1.6}{\percent} (\cref{fig:kinematic_variations}.c).
Overall, an increase in the maximum shape difference goes hand in hand with an increase of the relative tip displacement error.
The travelling wave index reconstruction shows excellent agreement over the entire range of tested kinematics (\cref{fig:kinematic_variations}.d and e).
The distribution of relative travelling wave index error (\cref{fig:kinematic_variations}.d) has a median of \qty{1.09}{\percent} and \qty{95}{\percent} of the data points are below \qty{3.03}{\percent}.
One data point has a large relative travelling wave index error, mainly due to the low reference value $\kindex{\mathcal{I}}{ref} = \num{3.4e-3}$.
The absolute error $|\mathcal{I} - \kindex{\mathcal{I}}{ref}|$ stays below \num{0.025} and has a median value of \num{1.3e-3}. 
These results demonstrate that the method is robust for various motions using clean data and the selected processing parameters: $\kindex{N}{j} = 10$ joints, an update interval $\tau/T = \num{5e-3}$, overlap parameter $\alpha = 0.6$ and a processing time increment equal to the update interval $\dd{t} = \tau$.
The effect of imperfect data and different choices of processing parameters are investigated next.

\subsection{Influence of hardware defects}
\begin{figure}
\centering
\includegraphics[width=\linewidth]{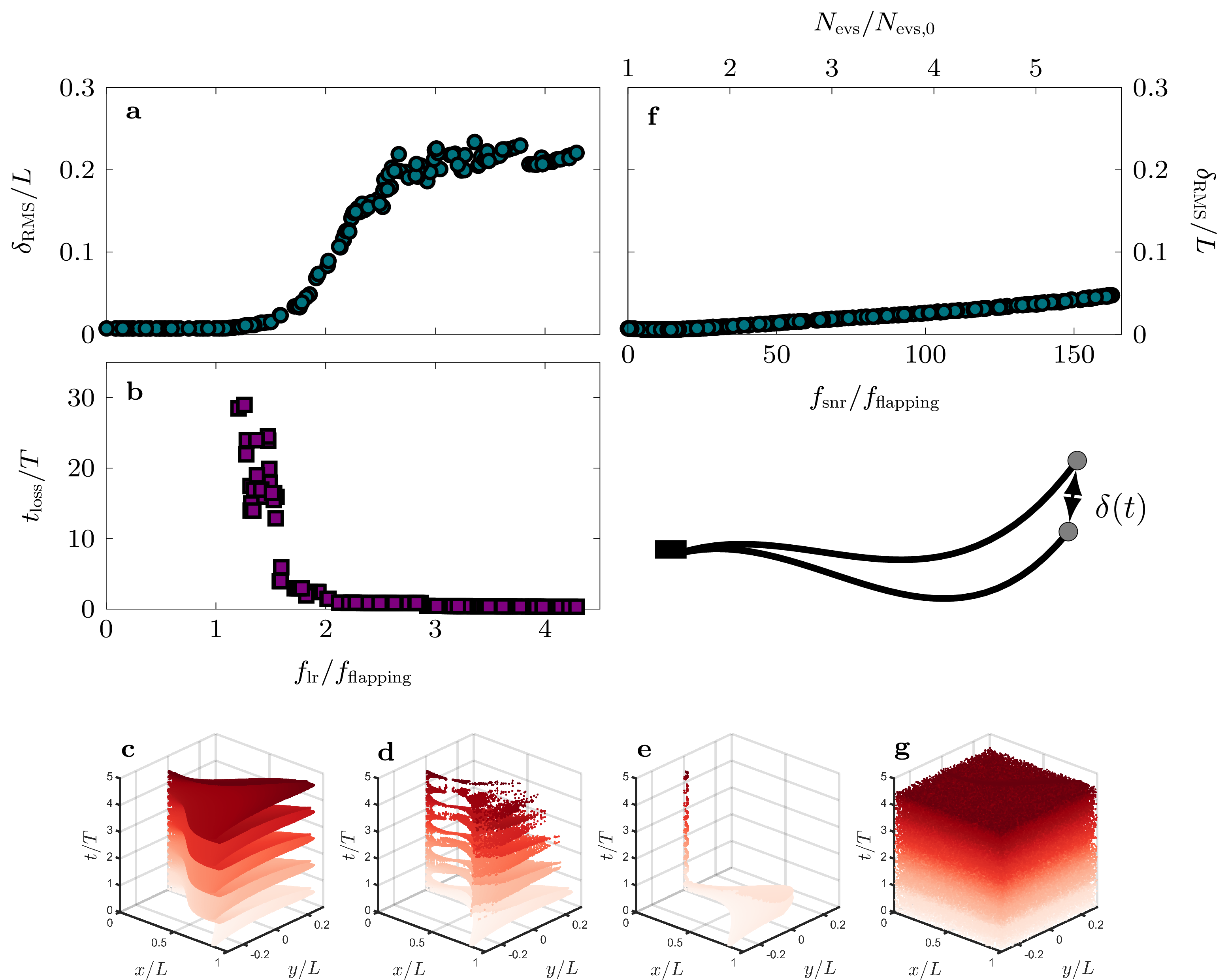}
\caption{Effect of simulated defects on the reconstruction accuracy: (a) root-mean square tip displacement error $\kindex{\delta}{RMS}/L$ as a function of the leak rate $\kindex{f}{lr}$. 
(b) The delay $\kindex{t}{loss}$ before the first occurrence of a tip displacement error greater than $0.1L$ is plotted in terms of number of cycles. 
(c, d and e) Visualisation of the leaked event clouds for  $\kindex{f}{lr} / \kindex{f}{flapping} = 0, 2.31, 3.89$. 
(f) Root-mean square tip displacement error $\kindex{\delta}{RMS}/L$ as a function of the dimensionless shot noise rate $\kindex{f}{snr} / \kindex{f}{flapping}$. 
The total amount of events in the raw file $\kindex{N}{evs}$ compared to the number of events in absence of shot noise $\kindex{N}{evs,0}$ increases linearly with the shot noise rate: at $\kindex{f}{snr} / \kindex{f}{flapping} = 37$, the amount of noise events equals the number of events describing the deformation ($\kindex{N}{evs} / \kindex{N}{evs,0} = 2$). 
(g) Visualisation of the noisy event clouds for $\kindex{f}{snr} / \kindex{f}{flapping} = 10$.}
\label{fig:defects_effect}
\end{figure}

Event-based cameras suffer from specific types of defects due to their particular hardware.
Electric leakage and temperature effects cause losses of the reference light intensity at the pixel level at a typical rate of $\kindex{f}{lr}$ \cite{nozaki_temperature_2017}.
Here, we tested the reconstruction accuracy and compared the results with the reference cases without event leakage ($\kindex{f}{lr} = 0$).
No loss of accuracy is observed for dimensionless leak rates $\kindex{f}{lr} / \kindex{f}{flapping} < 1.2$ as the tip displacement error remains the same as in the no leakage case (\cref{fig:defects_effect}.a).
For higher leakage rates $\kindex{f}{lr} / \kindex{f}{flapping} > 1.2$, the accuracy drops significantly.
The tip displacement error steps up from $\kindex{\delta}{RMS}/L = \qty{0.71}{\percent}$ to \qty{23}{\percent} when $\kindex{f}{lr} / \kindex{f}{flapping}$ increases from \qty{0} to $3.36$. 
The drop of accuracy originates from a severe loss of information in the cloud of events.
At $\kindex{f}{lr} / \kindex{f}{flapping} = 2.31$ (\cref{fig:defects_effect}.b), most of the events have disappeared after the fourth flapping cycle (\cref{fig:defects_effect}.c).
Deformations do not generate any event after the first half cycle for $\kindex{f}{lr} / \kindex{f}{flapping} = 3.89$ (\cref{fig:defects_effect}.d).

The delay of first occurrence $\kindex{t}{loss}$ of a large tip displacement error $\kindex{\delta}{RMS}/L > 0.1$ (\cref{fig:defects_effect}.a) describes how many cycles can be reconstructed accurately before the tip tracking is lost, and a reset is needed.
For low leak rate values $\kindex{f}{lr} / \kindex{f}{flapping} < 1.20$, notable tip displacement errors do not occur during the first \num{45} flapping cycles. 
When the leak ratio increases from \numrange{1.20}{1.60}, notable tip displacement errors are already identified after \num{3} flapping cycles.
For a dimensionless leak rate greater than \num{2}, the tip tracking is lost in less than one period.
Overall, the reconstruction is robust if the leak rate is lower than the characteristic flapping frequency ($\kindex{f}{lr} < \kindex{f}{flapping}$).
At room temperature, current sensors encounter $\kindex{f}{lr} \approx \qty{0.1}{\hertz}$ \cite{hu_v2e_2021}, yielding a dimensionless leak rate $\kindex{f}{lr} / \kindex{f}{flapping} \approx 0.01$ for flags flapping at $\qty{10}{\hertz}$, which is well below the threshold of accuracy loss.

Ghost events due to noise can occur due to photon flux variations and electronic noise, especially at low light intensities.
In the simulation, such ghost or shot noise events are random spurious events that are added to the actual stream of events that are triggered by an intensity change. 
They do not affect the occurrence or position of the actual events triggered by an intensity change.
The random injection of the shot noise events is governed in time by a Poisson process and in space by a random spatial distribution \cite{prophesee_metavision_2023}.
The average rate of occurrence of the shot noise events is indicated by $\kindex{f}{snr}$.
The total amount of events $\kindex{N}{evs}$ divided by the number of events in absence of shot noise $\kindex{N}{evs,0}$ is a relative measure of the added noise.
The shot noise rate \kindex{f}{snr} is normalised by the flapping frequency \kindex{f}{flapping} in \cref{fig:defects_effect}.e-f, to give an idea of the number of spurious events per flapping cycle.  
The tip displacement error stays below one percent for a dimensionless shot noise rate $\kindex{f}{snr}/\kindex{f}{flapping} <35$ (\cref{fig:defects_effect}.d).
At this critical noise rate, the desirable events triggered by deformation only represent \qty{51}{\percent} of the total stream, and \qty{49}{\percent} of the events come from noise ($\kindex{N}{evs} / \kindex{N}{evs,0} = 1.95$). 
This level of noise is extreme and the characteristic shape of the event cloud seen from experimental data in \cref{fig:defects_effect}.b is completely hidden in \cref{fig:defects_effect}.f where $\kindex{f}{snr}/\kindex{f}{flapping} = 10$. 

The order of magnitude of the shot noise rate in current sensors is found around $\kindex{f}{snr} \approx \qty{1}{\hertz}$ \cite{hu_v2e_2021} and can be reduced by cooling \cite{berthelon_effects_2018} and filtering \cite{ding_e-mlb_2023}.
This rate of shot noise is an order of magnitude lower that the typical flapping frequency of our flags which lies around $\qty{10}{\hertz}$. 
For $\kindex{f}{snr}/\kindex{f}{flapping} \approx 0.1$, the effect of shot noise is minimal. 

\subsection{Influence of processing parameters}
\begin{figure}
\centering
\includegraphics[width=\linewidth]{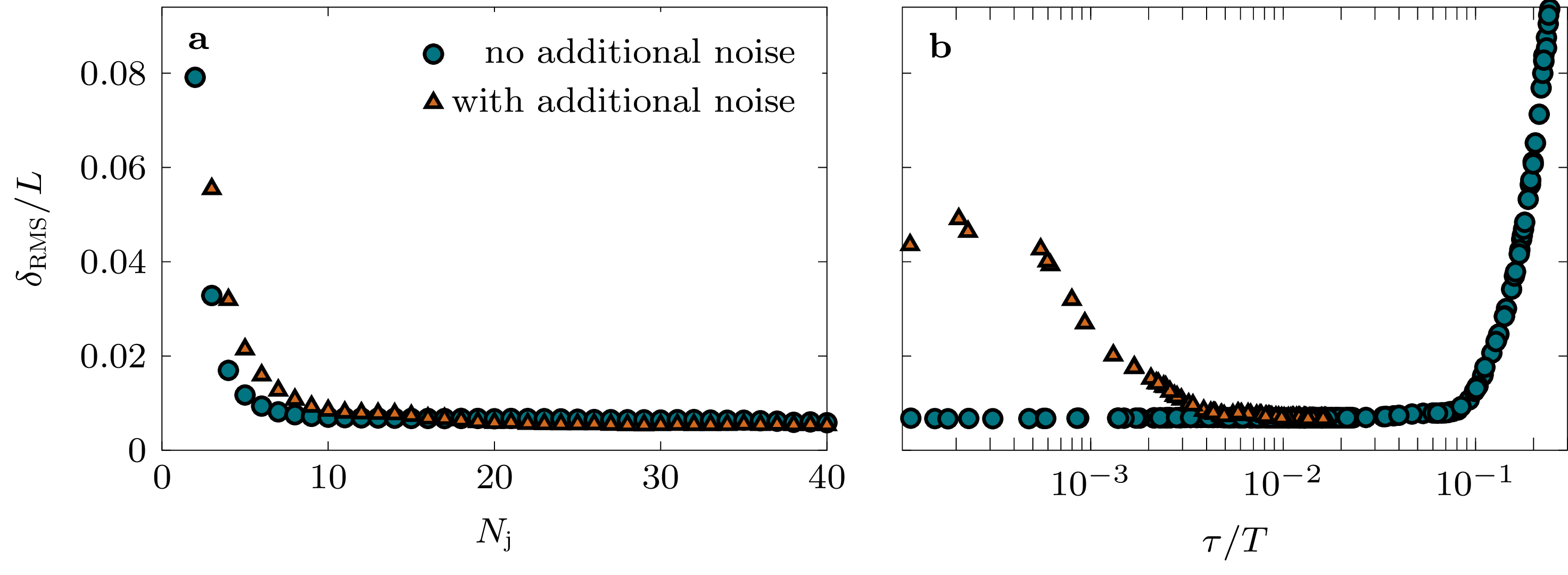}
\caption{Effect of (a) the number of joints $\kindex{N}{j}$ and (b) update interval $\tau$ on the root-mean square tip displacement error $\kindex{\delta}{RMS}$ with noiseless and noisy data. 
Shot noise level is set at $\kindex{f}{snr} / \kindex{f}{flapping} = 20$.}
\label{fig:ProcessingParameters}
\end{figure}

The effect of the processing parameters is illustrated with one kinematic of wavelength $\lambda/L=2$. 
Tip-based errors decrease rapidly with increasing number of joints up to $\kindex{N}{j} \approx 10$ (\cref{fig:ProcessingParameters}.a).
An increase in the joint numbers beyond $\kindex{N}{j} \approx 10$ does not yield a significant additional improvement of the approximation as the maximum curvature of the deformation is already well enough resolved by $\kindex{N}{j} \approx 10$ joints.
An analytical estimation of the number of joints based on the expected and observed curvature data is presented in the Appendix and confirms the results in (\cref{fig:ProcessingParameters}.a). 
More joints allow for smoother variations in curvature along the line but gather less events per joints.
The error stays low for larger number of joints.
The effect of the update interval on the tip displacement error is mild for $\num{2e-3} < \tau/T < \num{2e-2}$ (\cref{fig:ProcessingParameters}.b).
The error increases for lower and higher update intervals.
At low values ($\tau/T < \num{2e-3}$), the error increases with decreasing update interval for noisy data.
Reconstruction with noiseless data maintains small errors for very low update intervals ($\tau/T \approx \num{2e-4}$), where an average of less than 15 events are gathered per joint and update interval.
In this configuration, single events can locally modify joint coordinates resulting in an increased noise sensitivity.
At high values ($\tau/T > \num{2e-2}$), large update intervals act as a moving average and smooth out the motion.
Far away from these extreme cases, there exists a range of $\kindex{N}{j}$ and $\tau$ where the precision of the reconstruction is high and stable.

\section{Experimental results}
\label{sec:demo}

\begin{figure}
\centering
\includegraphics[width=\linewidth]{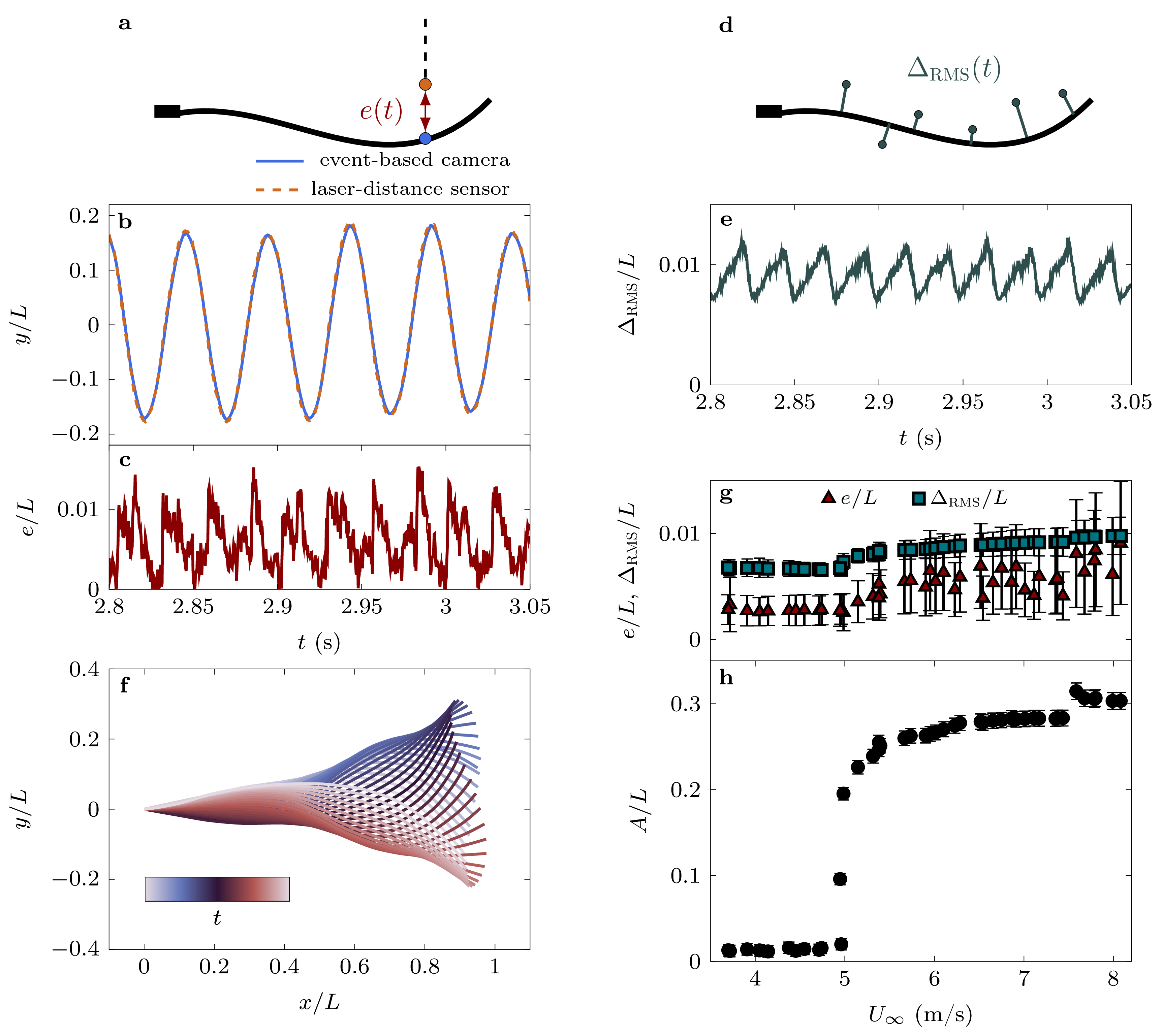}
\caption{Accuracy metrics for experimental data are computed in terms of 
(a) lateral displacement $y$ at a fixed $x$ position in the Cartesian or laboratory coordinate system using a laser distance sensor and 
(b) distance of the cloud of events to the reconstructed shape for each update interval $\kindex{\Delta}{RMS}$. 
For a $L =  \qty{18}{cm}$ flag at $\Uinf=\qty{7.2}{\meter\per\second}$ flapping at $f = \qty{20.7}{\hertz}$, 
(c) time-resolved comparison of the lateral displacement $y/L$ obtained at $x/L = 0.73$ with the event-based camera and the laser distance sensor. 
(d) The difference of both signals defines the error $e/L$. 
(e) Time variation of the cloud distance $\kindex{\Delta}{RMS}$ on the same time interval. (f) Series of reconstructed shapes coloured with time. 
(g) Time-average (marker) and standard deviation (error bars) for measurement time series of $\kindex{\Delta}{RMS}/L$ and $e/L$ at different inflow velocities \Uinf, and 
(h) time-average (marker) and standard deviation (error bars) of the tip amplitude $A/L$ for measurement series at different inflow velocities \Uinf.}
\label{fig:experimental_results}
\end{figure}

The proposed algorithm is now demonstrated for experimental data. 
A square flag made out of paper (Papeteria $\qty[mode = text]{80}{g.m^{-2}}$) of length $L =  \qty{18}{cm}$ is attached to a vertical pole at the outlet of an open-section wind tunnel (\qtyproduct[product-units=power]{450x450}{\mm}).
The centreline is illuminated by a horizontal light sheet generated with a continuous laser pointer and a Powell lens. 
An event-based camera (Century Arks SilkyEvCam VGA with a 35mm lens) is placed above the flag on one side and records the motion of the illuminated centreline for approximately \qty{6}{s}.
Reconstruction is performed with $\kindex{N}{j}=10$, $\alpha=0.6$, $\dd{s}=L/50$, $\dd{t} = \tau = \qty{5}{ms}$ for non flapping flags, and $\dd{t}=\tau = \qtyrange{0.29}{0.48}{ms}$ for flapping flags, resulting in an equivalent frame rate of $\qtyrange{2.1}{3.5}{kHz}$. 
A time-resolved laser distance sensor (Baumer OM70) measures independently the transverse displacement of the flag at $x/L = 0.73$.
Time synchronisation between the camera and the distance sensor is achieved with an external trigger signal. 
An absolute error $e(t)$ indicates the discrepancy between the distance sensor data and the reconstructed centreline from the event-based data (\cref{fig:experimental_results}.a).

Comparison between the lateral displacement data from the event-based camera and the distance sensor is presented in \cref{fig:experimental_results}.b.
The transverse error $e(t)$ oscillates between \qtyrange[range-phrase={\ and\ }]{0}{1.5}{\percent} of the centreline length (\cref{fig:experimental_results}.c).
The root-mean square distance $\kindex{\Delta}{RMS}$ between the cloud of points and the reconstruction surface for each update interval quantifies how close the reconstruction is from the original data (\cref{fig:experimental_results}.d).
The cloud of points is qualitatively well captured by the reconstruction surface with $\kindex{\Delta}{RMS}/L$ oscillating between \qtyrange[range-phrase={\ and\ }]{0.7}{1.2}{\percent} (\cref{fig:experimental_results}.e).
Superimposed deformations at different time instants outline the complex motion of the centreline and the two-dimensional path of the tip (\cref{fig:experimental_results}.f).
The envelope depicts two necks which is consistent with the literature for this type of paper flags and flow velocities \cite{virot_fluttering_2013}.

A series of deformation measurements is performed autonomously with the event-based camera (\cref{fig:experimental_results}.g and h).
The series of measurements go across several stages of fluid-structure interaction.
For $\Uinf < \qty{5}{\meter\per\second}$ the flag is stable and the amplitude $A/L \approx 0$.
For flow velocities ranging from $\qtyrange{5}{5.5}{\meter\per\second}$, the flapping amplitude first increases then plateaus before jumping up slightly at $\Uinf = \qty{7.5}{\meter\per\second}$ (\cref{fig:experimental_results}.h).
Our measurements are consistent with results from other experimental studies of flags that observe a first plateau of flapping amplitudes in the range $A/L \sim 0.24 - 0.34$ \cite{virot_fluttering_2013} and $A/L \sim 0.2 - 0.25$ \cite{eloy_origin_2012}.
Both the average distance of the reconstruction to the cloud of points and to the laser distance sensor data stay below $0.01L$ for each velocity (\cref{fig:experimental_results}.g).
The size of the raw event file at each velocity is a direct indicator of the flapping state of the flag because the number of events increases significantly for a flapping flag compared to a stable flag.
The file size allows real-time monitoring of the onset of flapping in the measurement loop that systematically increases the wind velocity without human supervision and without the need for any data processing.
Event-based cameras are thus ideally suited to be integrated into an automated experimental pipeline \cite{mulleners.2024}.
This set-up performs the measurement loop through different velocities in less than \qty{15}{\minute} for \num{44} measurement points, yielding around \qty{4.8}{GB} of raw data.
A high-speed camera recording 8 bit images of similar resolution (\qtyproduct{640x480}{px}) at 1000 frames per second would generate around \qty{80}{GB} of raw data, which is $16$ times more.

\section{Conclusion}

The interest of the research community in event-based cameras for fast imaging is increasing. 
The sensors and cameras have reached a level of maturity that makes them widely available at a price that is orders of magnitude lower than traditional high repetition rate scientific cameras. 
The difference in the data structure between event-based and frame-based imaging calls for the development of new processing algorithms that do not convert the streams of events back to snapshots, but instead leverage the full potential and information captured by the event detection.

Here, we presented a robust and generalisable method to reconstruct the spatio-temporal evolution of the centreline of a flapping flag from continuous raw streams of event data. 
Our algorithm relies on a coarse chain-like representation of the current state of the centreline and is updated by the occurrence of new events. 
The algorithm works with continuous illumination and leverages the full potential of event-based imaging to track fast deformations of objects tethered at a fixed location. 
The algorithm was applied to synthetic data and experimental data of flapping flags.
The influence of hardware defects and the processing parameters was evaluated. 
Common hardware defects will have negligible effects on the extraction of the flapping flag dynamics.
The choice of the processing parameters is a compromise between precision and robustness. 

The results from synthetic data and experimental campaigns reveal high levels of accuracy for various motions, wind conditions, and levels of defects.
Event-based imaging has the potential to unlock accurate tracking with high spatiotemporal resolution while reducing both storage needs and expenses by up to two orders of magnitude.
When integrated into autonomous measurement loops, event-based cameras enable the exploration of large parameter spaces and accelerate data-driven discoveries in experimental mechanics.
The combination of the low data rate of the cameras with our processing algorithm could even allow for live-monitoring or feedback control applications. 

\clearpage


\section*{Appendix}

\section*{Camera set up and calibration}

\begin{figure}
\centering
\includegraphics[width=\linewidth]{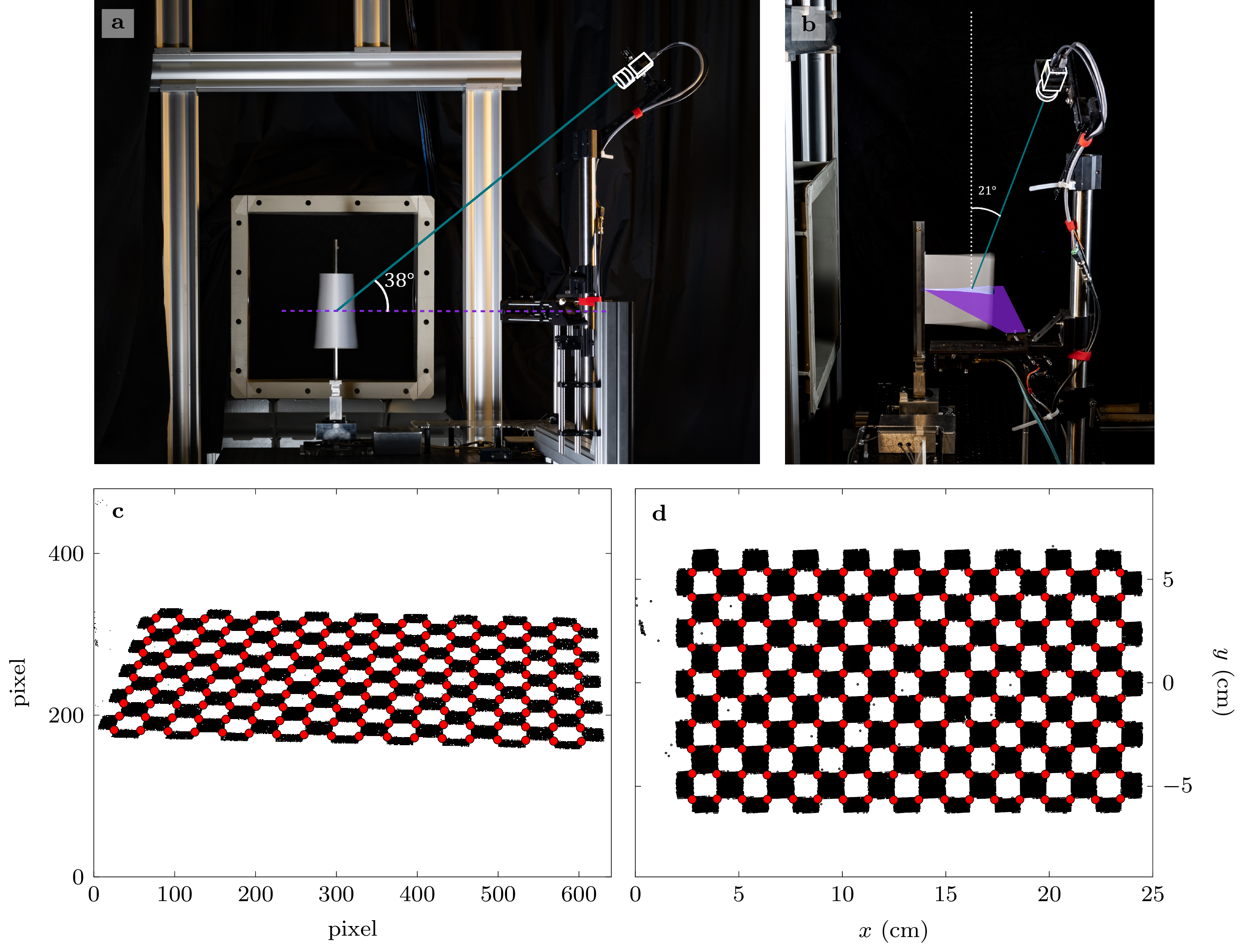}
\caption{The camera (white outline) is mounted next to outlet of the wind tunnel.
(a) On the front view of the set-up, the camera is tilted down with an angle of \ang{38} with respect to the light sheet plane (purple line).
(b) From the side view, the camera is oriented upstream with an angle of \ang{21}.
(c) A pseudo image of a blinking calibration pattern is created and the edges of the chequered board are detected (red dots).
(d) The calibration transforms the pixel coordinates to spatial coordinates.}
\label{fig:CameraAngle}
\end{figure}

The event-based camera is placed at the outlet of the wind tunnel and observes the intersection of the laser illumination plane and the flapping flag from above under an angle of \ang{38} (\cref{fig:CameraAngle}).
This orientation is compromise. 
If the camera would be placed directly above the flag (i.e. at an angle of $\ang{90}$ according to \cref{fig:CameraAngle}.a), the centreline can be hidden by the top part of the flag due to paralax effects and slight inhomogeneities in the deformation along the height of the flag.
If the camera would be placed under an angle of $\qty{0}{\deg}$, any depth perception is lost.
The lens (C-mount Computar $f=\qty{35}{mm}$ with a maximum aperture of $f/1.4$) is manually focused and the aperture is set at a value around $f/8$ to keep the moving centreline in focus.
The lens is positioned at an average working distance of the centreline of $d = \qty{1.02}{m}$, resulting in a magnification factor $M = f/(f-d) = 0.036$.
The $3/4$ inch sensor, with a total of \qtyproduct{640 x 480}{pixel}, has an effective physical resolution of $\eta = \qty{0.42}{mm}$.
Compared to the length of the flag, the dimensionless effective resolution is $\eta/L = \qty{0.23}{\percent}$.
A calibration of the camera accounts for the skewed field of view due to the tilted perspective.
A screen is positioned on the centreline plane and displays a checker board pattern.
The pattern is set to blink so that the event-based camera records intensity changes.
The recording is converted to a pseudo-image by accumulating events during one blinking period (\cref{fig:CameraAngle}.c).
The edge coordinates (red dots in \cref{fig:CameraAngle}.c and d) of the checker board are mapped to the coordinates of a regular grid pattern where $(x,y)=(0,0)$ corresponds to the root of the centreline.
The calibration mapping consists in a $\mathbb{R}^2 \rightarrow \mathbb{R}^2$ polynomial function taking the coordinates of the pseudo image space to the physical space of the centreline plane.
The resulting mapping has a residual maximum error of \qty{0.62}{mm} and an average error of \qty{0.18}{mm}.
During the centreline reconstruction, the calibration mapping is applied to each event in the stream so that the spatial clustering occurs in the physical space with no distortion.

\section*{Theoretical estimation of the required number of joints}

\begin{figure}
\centering
\includegraphics[width=\linewidth]{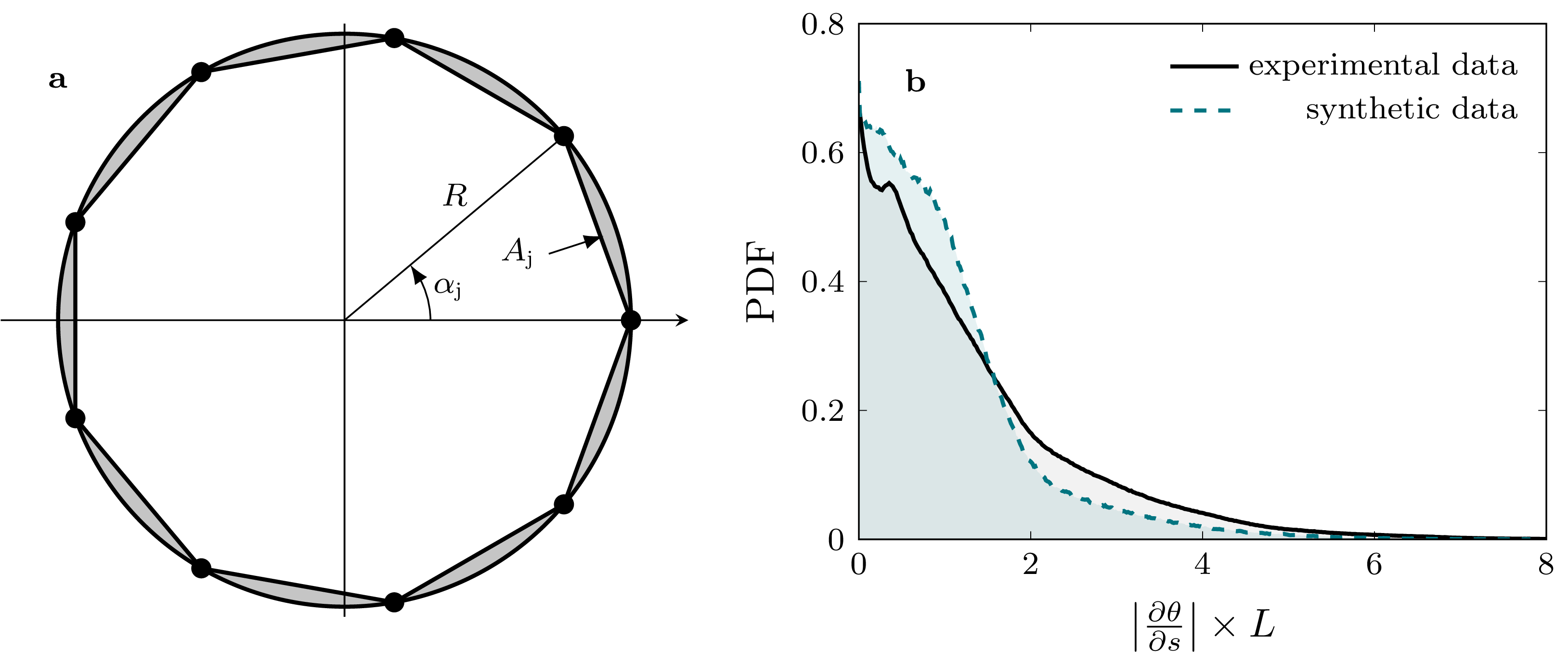}
\caption{(a) Circle with radius $R$ approximated by a chain with a finite number of $\kindex{N}{j}$ joints, equally distributed along the circle.
The area between the two curves is colored in gray.
(b) The probability density function from the histogram of the local curvature data gathered from the experimental measurement series and the synthetic data set.
}
	\label{fig:NumberOfJointsTheory}
\end{figure}

The precision of the approximation of a smooth curve by a set of straight links depends on the complexity of the shape and the density of joints.
We derive here a mathematical estimate of the number of joints $\kindex{N}{j}$ needed to reconstruct a deformation pattern with a given radius of curvature.
For a slowly varying curvature $1/R$, the problem is locally approximated as points distributed along a circle of constant radius of curvature $R$ (\cref{fig:NumberOfJointsTheory}.a).
The $\kindex{N}{j}$ joints divide the circle into triangular sections with an angle of $\kindex{\alpha}{j} = \frac{2\pi}{\kindex{N}{j}}$.
We consider first the error between the two curves based on the area between them (gray section in \cref{fig:NumberOfJointsTheory}.a).
The total area-based error is 
\begin{equation}
\kindex{N}{j} \kindex{A}{j} = \pi R^2 - \frac{1}{2}\kindex{N}{j}R^2 \sin\left(\frac{2\pi}{\kindex{N}{j}}\right)
\end{equation}
which behaves at first order as 
\begin{equation}
\kindex{N}{j} \kindex{A}{j} = \frac{2 \pi^3}{3} \left(\frac{R}{\kindex{N}{j}}\right)^2 + \mathcal{O} \left( \frac{1}{\kindex{N}{j}} \right)^4.
\end{equation}
We use the total length of the curve $L$ to make the area-based error dimensionless so that the shape error metric becomes:
\begin{equation}
\frac{\kindex{N}{j}\kindex{A}{j}}{L^2} \approx \frac{2 \pi^3}{3} \left(\frac{R}{L}\right)^2 \frac{1}{\kindex{N}{j}^2}.
\label{eq:Njoints}
\end{equation} 
This gives us an estimation of the required number of joints based on the maximum expected curvature of the deformation and acceptable shape error. 

To apply this estimate to our flapping flag data, we extract the distribution of the local radii of curvature for the experimental and synthetic data used for validation (\cref{fig:NumberOfJointsTheory}.b).
The upper bound of curvature from the experimental data is estimated at $L/R < 5.6$ from the 99\textsuperscript{th} percentile of all the concatenated data.
The curvature distribution from the synthetic data of different kinematics used in \cref{sec:bemchmark} and \cref{fig:kinematic_variations} has an upper bound of $L/R < 4.5$, which is of the same order of magnitude as in the experimental results.
With a target error of $\kindex{N}{j}\kindex{A}{j}/L^2 = \qty{1}{\percent}$ and an upper bound for the curvature of $L/R < 5.6$, we find a lower bound for the number of joints according to \cref{eq:Njoints}: 
\begin{equation}
\kindex{N}{j} \approx \sqrt{\frac{L^2}{\kindex{N}{j}\kindex{A}{j}}} \sqrt{\frac{2 \pi^3}{3}} \frac{R}{L} \geq 9.
\end{equation}
A similar analysis based on the approximation error in the curve length $\Delta L$ between the circle and its approximation using a chain with $\kindex{N}{j}$ joints yields the estimate:
\begin{equation}
\kindex{N}{j} \approx \sqrt{\frac{L}{\Delta L} \frac{\pi^3}{3} \frac{R}{L} } \geq 14
\end{equation}
for a target curve length error of $\Delta L/L = \qty{1}{\percent}$ and $L/R < 5.6$.
Both these estimates are conservative and we expect the smooth interpolation to reduce the area between the curves $\kindex{N}{j}\kindex{A}{J}$ and the length mismatch $\Delta L$.
Values of $\kindex{N}{j} \sim 10-15$ are consistent with our simulation results (\cref{fig:ProcessingParameters}.a) for different types of error metrics.

\section*{Supplementary information}

\textbf{Supplementary video S1:} example of a generated kinematics (reference of the motion is ms002mpt001 as described in the data set).\\
\textbf{Data and code availability:} The raw experimental data, the parameters and scripts to generate the simulation data, process the streams of events and post-process the results to obtain the figures are publicly available on Zenodo \url{https://doi.org/10.5281/zenodo.14290044}~\cite{Raynaud.2024.zenodo}.

\section*{Acknowledgments}

The authors wish to acknowledge the help from EPFL Center for Imaging through discussions with Dr. Edward Andò and his team.





\section*{References}

\bibliographystyle{iopart-num}
\bibliography{bibliography}

\end{document}